\documentclass[a4paper,twocolumn]{article}
\usepackage{graphics,overcite}
\title{The wormhole move: A new algorithm for polymer simulations}
\author{J. Houdayer\\
Institut f\"ur Physik, J. Gutenberg Universit\"at, D-55099 Mainz, Germany\\
Max Planck Institut f\"ur Polymerforschung, D-55021 Mainz, Germany\\
}
\begin{document}
\maketitle

\abstract{
A new Monte Carlo move for polymer simulations is presented. The
``wormhole'' move is build out of reptation steps and allows a polymer
to reptate through a hole in space; it is able to completely displace
a polymer in time $N^2$ (with $N$ the polymer length) even at high
density. This move can be used in a similar way as configurational
bias, in particular it allows grand canonical moves, it is applicable
to copolymers and can be extended to branched polymers. The main
advantage is speed since it is exponentially faster in $N$ than
configurational bias, but is also easier to program.}


\section{Introduction}
Polymer systems are very common (DNA, proteins, plastics, \ldots) and
theoretical studies are quite difficult. That is why efficient
numerical simulation techniques are required, either to test
theoretical approximations or to compare models with
experiments~\cite{Binder95}. Unfortunately polymer simulations are
very slow because they involve big time and length scales. A lot of
smart Monte Carlo moves have been devised to reduce the relaxation
time of such systems, in particular the pivot
algorithm~\cite{MadrasSokal88} for isolated polymer chains, the
reptation method (or ``slithering snake''
algorithm)~\cite{WallMandel75}, configurational bias~(CB)
\cite{SiepmannFrenkel92,FrenkelMooij92} and extensions around it
\cite{EscobedoPablo95b,ConstaWilding99}. Moreover general Monte Carlo
methods such as exchange Monte Carlo (or parallel
tempering)~\cite{HukushimaNemoto96}, histogram
re-weighting~\cite{FerrenbergSwendsen88} or ``go with the winner''
simulations~\cite{Grassberger97} are also used.

Here we propose a new kind of Monte Carlo move for polymer
simulations. This move, called ``wormhole'' (WH) move, has more or
less the same effect as CB but it is much faster since it is quadratic
in $N$ instead of exponential. Based on reptation moves, our algorithm
is relatively easy to program. Nevertheless our method is quite
different from standard reptation since it can do long-range
displacements and grand-canonical moves, it is also much more flexible
and can be applied to hetero and branched polymers.

In the next section, we precisely describe this new
method and prove its correctness, then we show with numerical
benchmarks that our algorithm is much faster than CB and finally we
introduce a set of possible modifications (in particular grand
canonical moves).

\section{The Algorithm}

\begin{figure}
\begin{center}
\resizebox{0.9\linewidth}{!}{\includegraphics{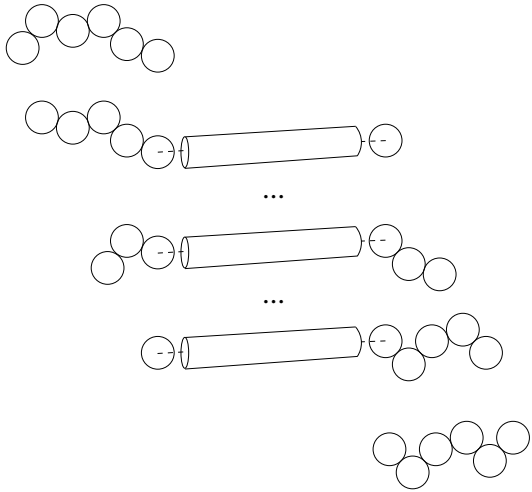}}
\end{center}
\caption{A {\it complete} WH move. Note how the monomers move as
they come and go through the hole.}
\label{wormhole_fig}
\end{figure}

We first roughly describe our algorithm. We consider a linear polymer
of length $N$ in a given fixed environment (temperature, other
polymers, solvent, \ldots). The WH Monte Carlo move opens a hole
in space (the wormhole) and tries to make the polymer reptate through
it (see figure~\ref{wormhole_fig}). The move is composed of a lot of
elementary reptation steps which are individually rejected or accepted
according to standard Metropolis. This goes on as long as the polymer
is not in one piece again. Each of the reptation step is randomly
backward or forward and the polymer thus does a discrete
one-dimensional random walk. Three different scenarios may happen:
\begin{enumerate}
\item The very first step fails, the hole is not opened and nothing happens.
We call it a {\it failed} move.
\item The polymer reptates through the hole as sketched in
figure~\ref{wormhole_fig}; the random walk has covered a distance
$N$. The polymer is in a completely new location and new
configuration. This is a {\it complete} move.
\item The polymer enters the hole but comes out on the
same side; the random walk got back to its starting point. The
monomers which got through the hole and came back have moved and the
others are unchanged. This is a {\it partial} move.
\end{enumerate}

Let us now enter into the details and describe exactly what the
algorithm does. WH proceeds as follow:
\begin{enumerate}

\item Wormhole drilling step: Randomly choose one end-monomer (with
probability $1/2$) and try to move it to a uniformly random position,
with the old bond broken and a virtual one created to the other end of
the polymer (the dashed line in figure~\ref{wormhole_fig}). Proceed to
step 3.

\item Standard reptation step: Randomly choose one end-monomer and try to
move it to the other end of the polymer connecting it with a uniformly
random bond. For this move, we do as if the virtual bond was a normal
one, so that the polymer has only two ends.

\item End test: If the polymer is in two pieces, proceed to step 2.
Otherwise the WH move is finished and is accepted with probability
one.

\end{enumerate}
At step 1 and 2, the move is accepted with the Metropolis probability:
\begin{equation}
P_M(\Delta E) = \min\left(1,e^{-\beta\Delta E}\right),
\label{metro_eq}
\end{equation}
where $\beta=1/k_B T$ and $\Delta E$ is the energy difference
generated by the trial move. Note that the virtual bond may be given a
constant energy to improve the acceptance rate of the first step.

The claim is that WH is a correct Monte Carlo move: it overall
respects the detailed balance condition:
\begin{equation}
P(C)P(C\rightarrow C')=P(C')P(C'\rightarrow C),
\label{det_bal_eq}
\end{equation}
where $P(C\rightarrow C')$ is the transition probability from
configuration $C$ to $C'$ and $P(C)=e^{-\beta E(C)}/Z$ is the
Boltzmann weight. It is easy to check that the new configuration is a
valid one~: (i) there is neither hole nor virtual bond any longer,
(ii) the nature of the monomers and their order along the chain are
the same as in the old configuration (that is why it works for
hetero-polymers, in contrast to simple reptation).

If the first step is rejected, equation~\ref{det_bal_eq} is trivially
fulfilled. Otherwise, $n$ elementary steps are performed and the
system goes through a sequence of configurations $C_0,\cdots,C_n$ with
energies $E_0,\cdots,E_n$. Each elementary step
$S_i=C_{i-1}\rightarrow C_i$ takes the system from one configuration
to the next (with $C_{i-1}=C_i$ if the step is rejected). The
probability of the sequence $S=(S_1,\cdots,S_n)$ can be written
\begin{equation}
P(S) = \prod_{i=1}^n P(S_i).
\end{equation}
To each sequence corresponds an opposite sequence
$\overline{S}=(\overline{S_n},\cdots,\overline{S_1})$, made of the
opposite steps $\overline{S_i}=C_i\rightarrow C_{i-1}$, which takes
the system from $C_n$ to $C_0$ and we have
\begin{equation}
\frac{P(S)}{P(\overline{S})} = \prod_{i=1}^n
\frac{P(S_i)}{P(\overline{S_i})}.
\end{equation}
For all accepted intermediate steps ($1<i<n$), we have:
\begin{eqnarray}
\nonumber\frac{P(S_i)}{P(\overline{S_i})} & = &
\frac{\frac12 P_B(S_i)P_M(E_i-E_{i-1})}
{\frac12 P_B(\overline{S_i})P_M(E_{i-1}-E_i)}\\ & = &
e^{-\beta(E_i-E_{i-1})},
\label{step_eq}
\end{eqnarray}
where $P_B$ is the uniform probability of choosing a bond and the
factor $1/2$ comes from the choice of the reptation direction. The
rejected steps fulfil $S_i=\overline{S_i}$ and cancel. They thus also
fulfil equation~\ref{step_eq}. On the other hand, for $i=1$,
\begin{equation}
\frac{P(S_1)}{P(\overline{S_1})}=\frac{P_P(S_1)}{P_B(\overline{S_1})}
e^{-\beta(E_1-E_0)},
\end{equation}
and for $i=n$,
\begin{equation}
\frac{P(S_n)}{P(\overline{S_n})}=\frac{P_B(S_n)}{P_P(\overline{S_n})}
e^{-\beta(E_n-E_{n-1})},
\end{equation}
with $P_P$ the uniform probability of choosing a position within the
simulation box. Hence we get
\begin{equation}
\frac{P(S)}{P(\overline{S})} = e^{-\beta(E_n-E_0)}.
\label{seq_eq}
\end{equation}
Since $P(C\rightarrow C')$ is the sum of $P(S)$ over all possible
sequence $S$ with $C_0=C$ and $C_n=C'$, we get
equation~\ref{det_bal_eq}. This proves the correctness of our
algorithm.

Finally, concerning ergodicity, it is clear that this algorithm is
able to move a polymer to any place where the required space exists
beforehand. But, as for reptation and CB, one cannot for example
ensure global ergodicity on a very densely occupied
lattice. Nevertheless, as shown in the next section, WH appears to
perform quite well even at relatively high density.

\section{Efficiency}
Now that the algorithm is proven correct, let us see how good it
is. Usually the main difficulty for a Monte Carlo algorithm for
polymer is to relax the configuration of the polymers, which is more
an entropy problem than an energy problem. That is why we test the
speed of this new algorithm on the bond fluctuation model (BFM)
\cite{CarmesinKremer88,DeutschBinder91} at infinite temperature. The
BFM is a lattice polymer model where each monomer occupies 8 sites
($2\times 2\times 2$) of a cubic lattice and monomers cannot overlap
(hard core repulsion). The bond vectors are restricted to a set of 108
possibilities with lengths $2,\sqrt{3},\sqrt{5},3$ and
$\sqrt{10}$. The interaction range is set to $\sqrt{6}$ which means
that only monomers in direct contact interact.

\begin{figure}
\begin{center}
\resizebox{0.9\linewidth}{!}{\includegraphics{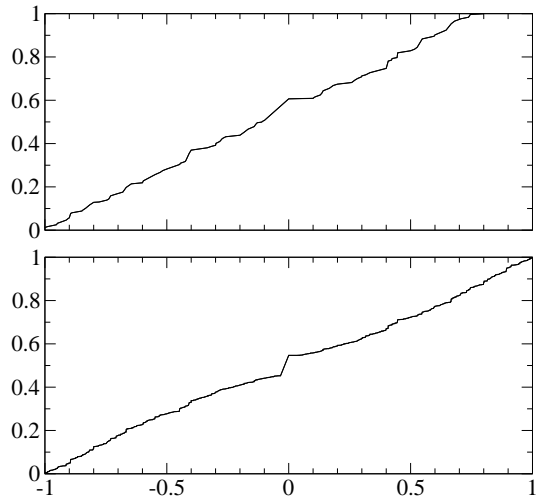}}
\end{center}
\caption{
Comparison of the conformations produced by WH and by standard
reptation for an isolated polymer of size $N=50$ at $T=2$. Both
figures show two curves superimposed, but they are
indistinguishable. Top: the integrated distributions of $\cos\theta$
with $\theta$ the angle between two successive bonds. Bottom: the
integrated distributions of $\sin\phi$ with $\phi$ the (torsion) angle
between a bond and the plane defined by the two preceeding bonds. }
\label{angles_fig}
\end{figure}

To check the correctness of the implementation, we used our program to
simulate isolated chains of different lengths with attractive
interactions near the $\Theta$-temperature (namely $T=2$). This is the
most sensitive point to check that the conformation of the chain is
correctly sampled, since it is where the fluctuations and the
temperature dependence are the largest. First we reproduced the
measurements of the $\Theta$-temperature done by Wilding {\it et
al.}~\cite{WildingMueller96} and found fully compatible results
(namely $T_\Theta=2.03(3)$ to be compared with
$T_\Theta=2.02(2)$). Moreover we compared the distributions of the
bond-bond and torsion angles with the ones given by standard
reptation. The results are shown on figure~\ref{angles_fig}. Both
algorithms give the same results (within statistical error bars, here
roughly $0.1\%$) and the curves, on top of one another, are
indistinguishable. Note that the roughness of the curves is no noise,
it is due to the discreteness of the model.

\begin{figure}
\begin{center}
\resizebox{0.9\linewidth}{!}{\includegraphics{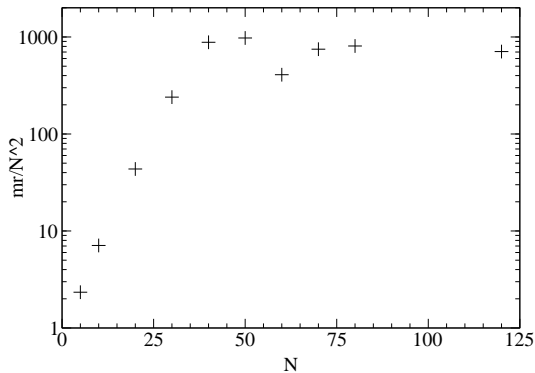}}
\end{center}
\caption{The number of accepted elementary steps required to achieve
a {\it complete} WH move is $O(N^2)$.}
\label{n2_fig}
\end{figure}

All the following benchmarks are made using a mixture of polymers of
length $N$ at density $1/2$ in the BFM at infinite temperature. For
these parameters, our computer (a Pentium III at 500 MHz) achieve $1.3\
10^6$ elementary steps per second with an acceptance rate $r\simeq
8.5\%$. Note that our implementation also works at finite temperature
and is thus slower than an implementation optimised for the infinite
temperature case.

First, we have claimed that our algorithm is able to completely
displace a polymer in time $N^2$. This is a reasonable assumption
since it is the time required by an unbiased random walk to cover a
distance $N$ and this corresponds to a {\it complete} WH move. But
this must be tested numerically since the random walk actually done is
strongly influenced by density fluctuations. We note $m$ the mean
number of elementary steps required to achieve one {\it complete} WH
move (many {\it failed} and {\it partial} moves may also happen in
between) and $mr$ is the corresponding number of accepted elementary
steps. The values of $mr/N^2$ are shown as a function of $N$ in
figure~\ref{n2_fig}, they clearly show that for large $N$, we have
\begin{equation}
m=\frac{\alpha}{r} N^2,
\label{n2_eq}
\end{equation}
with $\alpha\simeq 800$. To achieve a {\it complete} move on our
computer, we thus need around $7.\ 10^{-3}N^2$ seconds for a long
polymer (and less for a smaller one). As could be expected, the value
of $\alpha$ decreases at lower density, for example at density $0.1$,
$\alpha\simeq 1.8$ (and $r\simeq67.5\%$).

\begin{figure}
\begin{center}
\resizebox{0.9\linewidth}{!}{\includegraphics{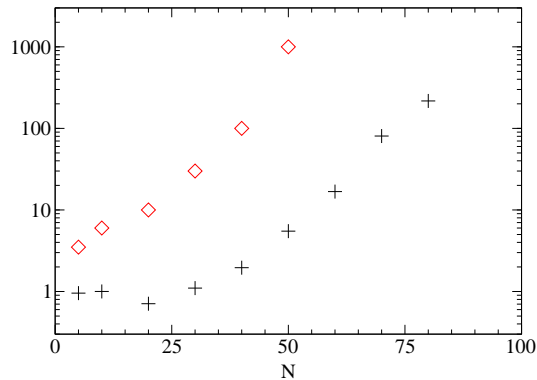}}
\end{center}
\caption{CPU time ratios between CB and WH as a function of $N$.
Crosses are time ratios for completely displacing a polymer (time to
accept a CB move/time to do a {\it complete} WH move). Diamonds are
relaxation time ratios (see equation~\ref{relax_eq}).}
\label{cb_fig}
\end{figure}

What our algorithm can do is very similar to CB, so it is interesting
to do the comparison. The CB with which we compare tries to erase a
polymer completely and to recreate it somewhere else, for each monomer
it checks the 108 possible bonds and chooses an allowed one (not
already occupied) and computes the associated weight. On our computer,
we can do $7. 10^4$ such elementary steps per second (one needs $N$ of
them for a whole CB move). As for WH, our CB implementation
also works at finite temperature and is thus slower than it should for
the infinite temperature case. Note that one could also use a CB
implementation which does not check all the 108 bonds (say 10 or
20). In this case the elementary steps are accordingly faster but the
overall acceptance rate decreases much faster with $N$.

In figure~\ref{cb_fig}, we show the ratios of time required by both
algorithm to completely displace a polymer (an accepted CB move and a
{\it complete} WH move). From this point of view, the algorithms are
more or less equivalent at small $N$, and WH becomes exponentially
better at large $N$. In this first comparison we did not take into
account the fact that WH also relaxes the system when it does a {\it
partial} move. So we measured the relaxation time of the end-to-end
vector $\mathbf{R}(t)$, namely:
\begin{equation}
\tau=\frac1{\langle \overline{\mathbf{R}^2} \rangle}
{\int\langle \mathbf{R}(0)\cdot\mathbf{R}(t) \rangle \mbox{d}t}.
\label{relax_eq}
\end{equation}
The ratios of those times are also shown on figure~\ref{cb_fig},
clearly WH is much faster than CB. On the other hand, at a lower
density, CB is more efficient for the small values of $N$ (for example
at density 0.1 and at $N=40$, CB is only 3 times slower than WH
instead of 100 times slower), but CB stays exponentially slower at
large $N$. See below how one could take advantage of a faster CB for
small $N$.

Finally this comparison was done on a lattice, this gives a big
advantage to CB since off-lattice CB implementations require a large
number of energy computations which are much longer than on a
lattice. Moreover, reptation is hampered by the lattice since it can
get blocked much more easily than off-lattice. We thus think that WH
should be even better for continuous models.

Our WH method is reminiscent of the extended ensemble method
introduced by Escobedo and de Pablo
\cite{EscobedoPablo95,EscobedoPablo96}. But in their method, the
elementary steps are CB moves that move $k$ monomers at a
time. Moreover the intermediate configurations are also sampled giving
rise to extended ensembles with configurations where a polymer is
incomplete or split into two pieces. Since they relax the whole system
between each elementary step the time to deal with one polymer is
\begin{equation}
m'=\frac{\alpha'V}{r'}\left(\frac Nk\right)^2,
\end{equation}
with $V$ the number of particles in the system and $\alpha'$ and $r'$
the equivalent of $\alpha$ and $r$ in equation~\ref{n2_eq}. Here
$r'\propto \exp(- k/k_0)$ and there is an optimal value for $k$ (namely
$k=2k_0$). With our parameters $k_0\simeq 4.5$ and $r'\simeq 3\%$, so
even if $\alpha'<\alpha$, $V/k^2$ is very large.

Finally, WH has one feature which may be undesirable. The execution
time for one move has huge fluctuations (that explains the
fluctuations in figure~\ref{n2_fig}). A move can be very fast (for
example a {\it failed} move) but it can also last a few seconds (and
exceptionally a few minutes for long chains). In a simple simulation
that is no problem, since only the average time is relevant. But in a
parallel computation, where each processor waits after the slowest
one, this can become a real difficulty. A first solution is to do a
lot of WH moves between two synchronisations, thus averaging the
fluctuations. Another one is to set a limit to the number of
elementary steps which can be done in one move (say $10 m$). If this
threshold is exceeded, the move is rejected and the polymer is
restored to its initial state.

\section{Variations}

Let us first see algorithmic refinements to WH. First, in the case of
an off-lattice simulation one may wish to draw the polymer bonds out
of a non-uniform $P_B$ distribution. In particular one may want to
draw the bond length around a certain favourable value. In such a case
it is necessary to add a correcting weight to the Metropolis rule to
recover equation~\ref{step_eq} (as done for normal reptation). Then
one can check that these correcting weights raise the correct overall
correction term and equation~\ref{seq_eq} still holds so the
algorithm is still correct. More generally, as soon as one wants to
modify the WH algorithm, one should check that the cancellations still
occur, in particular for $i=1$ and $i=n$.

In the original reptation method~\cite{WallMandel75} the direction of
the reptation is changed only when a move is rejected and not drawn
randomly at each step as we propose here. Nothing prevents to do the
same in WH (the proof just gets a little more complicated). At very
low density this can improve the speed by a factor 2 or more because
successive steps are done in the same direction. But at high density,
it changes essentially nothing.

As explained in the previous section, CB can be more efficient at
small $N$ than WH (at low density or when we do not check all the 108
bonds at each step). Then, as done in the ``extended ensemble''
method, one can use CB as the elementary step, moving $k$ monomers at
a time and we then have (with notation of the previous section)
\begin{equation}
m=\frac{\alpha}{r'}\left(\frac Nk\right)^2,
\end{equation}
and the CPU time is proportional to $km$ (the optimal value of $k$ is
now $k=k_0$). This is faster than the original WH if $kr'/r>K$ with
$K$ the CPU time ratio to do one elementary move with both methods
(here $K\simeq 19$). Note that we have supposed that $\alpha$ does not
change which is not clear. Finally one of the interesting point in WH
is the ease of programming, this is lost in this variation.

Let us now see how WH can be extended. First, how to do grand
canonical moves. These moves can be of three different kinds: (i) to
move a polymer from one simulation box to another one; (ii) to remove
one polymer from a box; (iii) to insert a new polymer in a box. The
first one is very easy to do, WH applies directly and one can
eventually add a chemical potential for each monomer depending on the
box. For (ii) and (iii), one proceeds more or less as for CB. The
algorithm has to be modified, the reptation steps are made between the
simulation box and some kind of black box without geometry where the
monomer have a given chemical potential. The $P_B$ and $P_P$ no longer
completely cancel (the steps toward the black box do not require a
random draw) and for an inserting move and the corresponding removing
move, equation~\ref{seq_eq} becomes
\begin{equation}
\frac{P(S)}{P(\overline{S})} = e^{-\beta(E_n-E_0)} P_P P_B^{N-1}.
\end{equation}
The new factor is a constant which can be absorbed in the chemical
potential (exactly as the weight with CB). This factor is exactly the
probability for choosing a particular configuration of the polymer. We
did not perform tests for the grand canonical moves, but we expect
them to behave as described in the previous section.

\begin{figure}
\begin{center}
\resizebox{0.9\linewidth}{!}{\includegraphics{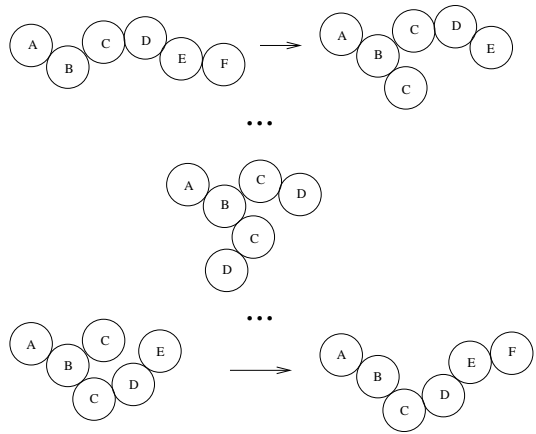}}
\end{center}
\caption{A {\it complete} modified WH move. The letters represent
different kind of monomers in a hetero-polymer. Note how in the
intermediate configurations, monomer F has disappeared and monomer C
is present twice.}
\label{modified_fig}
\end{figure}

Very often CB is not used to remove and insert a whole polymer, but is
applied only on a part of it. Such a thing can also be done with WH,
provided we perform a little modification (see
figure~\ref{modified_fig}). At the first step of the algorithm,
instead of choosing a random position, the monomer is randomly bound to
another non-end monomer of the same polymer. After this, WH proceeds
as before until the polymer is linear again. Note that in the case of
a hetero-polymer, the monomer that goes through the hole may be
replaced by another one, so that the polymer has kept its structure at
the end (see figure~\ref{modified_fig}). This modified move is
probably less efficient than the original one for the same number of
monomers, since the new growing branch is going to encounter an
already densely occupied region (the old branch being also
there). Moreover the original WH is also able to move parts of the
polymer (with {\it partial} moves), so it is perhaps better to use the
original move. The modified move can in any case be useful for
branched polymers.

\begin{figure}
\begin{center}
\resizebox{0.9\linewidth}{!}{\includegraphics{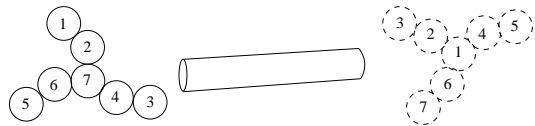}}
\end{center}
\caption{The order in which the monomers enter the hole (left), and leave it
(right) for a branched polymer.}
\label{star_fig}
\end{figure}

Finally one can also extend the original WH to branched polymers. The
basic idea stays the same: a random walk through a set of non-valid
configurations where the polymer is split into two pieces. But it
cannot use simple reptation because of the branching points. One way
to achieve this is to make the monomers get out of the hole in a
different order as they enter it. See figure~\ref{star_fig} for an
example. As in the previous paragraph, during the move, certain
monomers may be present twice whereas others are no longer there.

\section{Discussion and Conclusions}

The wormhole move (WH) presented here is a new kind of Monte Carlo
move for polymer simulations. It can completely displace a polymer in
time $O(N^2)$ even at high density ($1/2$ in the bond fluctuation
model). It is based on reptation steps and is thus quite easy to
program. It is nevertheless very flexible since it works on
hetero-polymers, allows grand canonical moves (insertions and
deletions of polymers) and can be adapted to branched polymers.

What is this new method good for ? Essentially, it is meant to replace
configurational bias (CB). The wormhole move can in fact do whatever
CB does, and it does it faster. As CB, it can destroy, create or
displace a polymer, but CB works in time $O(e^{\gamma N})$ whereas WH works
in time $O(N^2)$. Thus, exception made of the case of short polymers
at low density, it is always better to use WH instead of CB. In the
case of short polymers at low density, it is probably better to use WH
as well, since this case is anyway an easy one which requires short
computation time and WH is relatively easier to program.

On the other hand, the aim of WH is not to replace the standard
reptation method or to compete with it. It simply does something
different. Typically standard reptation is to be preferred in the case
of linear homo-polymers with small density fluctuations (i.e. when it
is not important to be able to rapidly displace a polymer between two
distant positions). In the other cases, where one would have used CB,
one should use WH. There are essentially two main cases: (i) if
reptation is not applicable, in particular to do grand canonical moves
or to simulate hetero-polymers; (ii) in the case of large density
fluctuations (typically when more than one phase is present)
where one wants to rapidly relax the density by displacing polymers
over large distances.

We expect this new method to become widely used since it is easy to
program and very efficient. At the moment, it is used to simulate
dense random copolymer mixtures~\cite{HoudayerMueller}.

The author acknowledges M. M\"uller, L. G. MacDowell, A. Yethiraj and
K. Binder for fruitful discussions.

\makeatletter \renewcommand\@biblabel[1]{$^{#1}$} \makeatother
\bibliographystyle{jcp}
\bibliography{/remote/home/plato/houdayer/Papers/Biblio/references}

\addcontentsline{toc}{chapter}{\protect\bibname}
\begin{thebibliography}{10}

\bibitem{Binder95}
{\em {M}onte {C}arlo and Molecular Dynamics Simulations in Polymer Science}, K.
  Binder, ed., (Oxford University Press, New York, 1995).

\bibitem{MadrasSokal88}
N. Madras and A. Sokal, J. Stat. Phys. {\bf 50,} 109 (1988).

\bibitem{WallMandel75}
F.~T. Wall and F. Mandel, J. Chem. Phys. {\bf 63,} 4592--4595 (1975).

\bibitem{SiepmannFrenkel92}
J.~I. Siepmann and D. Frenkel, Mol. Phys. {\bf 75,} 59--70 (1992).

\bibitem{FrenkelMooij92}
D. Frenkel, G.~C. M.~A. Mooij, and B. Smit, J. Phys.: Condens. Matter {\bf 4,}
  3053--3076 (1992).

\bibitem{EscobedoPablo95b}
F.~A. Escobedo and J.~J. {de Pablo}, J. Chem. Phys. {\bf 102,} 2636--2652
  (1995).

\bibitem{ConstaWilding99}
S. Consta, N.~B. Wilding, D. Frenkel, and Z. Alexandrowicz, J. Chem. Phys. {\bf
  110,} 3220--3228 (1999).

\bibitem{HukushimaNemoto96}
K. Hukushima and K. Nemoto, J. Phys. Soc. Jpn. {\bf 65,} 1604--1608 (1996).

\bibitem{FerrenbergSwendsen88}
A.~M. Ferrenberg and R.~H. Swendsen, Phys. Rev. Lett. {\bf 61,} 2635--2638
  (1988).

\bibitem{Grassberger97}
P. Grassberger, Phys. Rev. E {\bf 56,} 3682--3693 (1997).

\bibitem{CarmesinKremer88}
I. Carmesin and K. Kremer, Macromolecules {\bf 21,} 2819--2823 (1988).

\bibitem{DeutschBinder91}
H.-P. Deutsch and K. Binder, J. Chem. Phys. {\bf 94,} 2294--2304 (1991).

\bibitem{WildingMueller96}
N.~B. Wilding, M. M{\"u}ller, and K. Binder, J. Chem. Phys. {\bf 105,} 802--809
  (1996).

\bibitem{EscobedoPablo95}
F.~A. Escobedo and J.~J. {de Pablo}, J. Chem. Phys. {\bf 103,} 2703--2710
  (1995).

\bibitem{EscobedoPablo96}
F.~A. Escobedo and J.~J. {de Pablo}, J. Chem. Phys. {\bf 105,} 4391--4394
  (1996).

\bibitem{HoudayerMueller}
J. Houdayer and M. M{\"u}ller, in preparation.

\end{thebibliography}
\end{document}